\documentclass[rnote,traditabstract]{aa} 
\usepackage{graphicx}
\usepackage{amsmath} 
\usepackage{amssymb}
\usepackage{lscape}
\usepackage{natbib}
\usepackage{pifont}

\bibpunct{(}{)}{;}{a}{}{,}

\usepackage{txfonts}
\begin{document}

   \title{VLT/X-shooter Spectroscopy of a Dusty Planetary Nebula
     Discovered with Spitzer/IRS}

   \author{Isa Oliveira\inst{1}, 
           Roderik A. Overzier\inst{2}, 
           Klaus M. Pontoppidan\inst{3}, 
           Ewine F. van Dishoeck\inst{1,4}
          \and
           Loredana Spezzi\inst{5}
          }

   \institute{Leiden Observatory, Leiden University, P.O. Box 9513,
     2300 RA Leiden, The
     Netherlands\\ \email{oliveira@strw.leidenuniv.nl} \and
     Max-Planck-Institut f\"ur Astrophysik, Karl-Schwarschild-Str. 1,
     D-85741 Garching, Germany \and California Institute of
     Technology, Division for Geological and Planetary Sciences, MS
     150-21, Pasadena, CA 91125, USA \and Max-Planck Institut f\"ur
     Extraterrestrische Physik, Giessenbachstrasse 1, 85748 Garching,
     Germany \and Research and Scientific Support Department, European
     Space Agency (ESA-ESTEC), P.O. Box 299, 2200 AG Noordwijk, The
     Netherlands
}

   \date{Received ; accepted }

 
  \abstract {As part of a mid-infrared spectroscopic survey of young
    stars with the {\it Spitzer Space Telescope}, an unclassified red
    emission line object was discovered. Based on its high ionization
    state indicated by the {\it Spitzer} spectrum, this object could
    either be a dusty Supernova Remnant (SNR) or a Planetary Nebula
    (PN). In this research note, the object is classified and the
    available spectroscopic data are presented to the community for
    further analysis. UV/optical/NIR spectra were obtained during the
    science verification run of the VLT/X-shooter. A large number of
    emission lines are identified allowing the determination of the
    nature of this object. The presence of strong, narrow ($\Delta v$
    $\sim$8 -- 74 km s$^{-1}$) emission lines, combined with very low
    line ratios of, e.g., [N {\sc ii}]/H$\alpha$ and [S {\sc
        ii}]/H$\alpha$ show that the object is a Planetary Nebula (PN)
    that lies at an undetermined distance behind the Serpens Molecular
    Cloud. This illustrates the potential of X-shooter as an efficient
    tool for constraining the nature of faint sources with unknown
    spectral properties or colors.}


\authorrunning{Oliveira et al.}
\titlerunning{Discovery of a Planetary Nebula}

   \maketitle
%

\section{Introduction}

As part of a program with the InfraRed Spectrograph (IRS,
\citealt{HO04}) onboard the {\it Spitzer Space Telescope} (SST,
\citealt{WE04}) aimed at characterizing the circumstellar disks of a
flux-limited sample of infrared-excess young stellar objects (YSOs) in
the Serpens Molecular Cloud ($\alpha_{J2000}$=18$^h$29$^m$49$^s$,
$\delta_{J2000}$=+01$^d$14$^m$48$^s$), an interesting object of
unknown nature was discovered (SSTc2dJ18282720+0044450, or \#17 in
\citealt{OL10}, hereafter OL17). The very bright, high ionization
emission lines seen in the mid-IR spectrum of this object are not
consistent with it being a YSO.

Two types of galactic objects show such high excitation lines: dusty
planetary nebulae (PNe, \citealt{BS09,ST07,GU07}) and supernova
remnants (SNRs, \citealt{SA09,GA09}). SNRs typically show very broad
emission lines, produced by the high velocity shock waves
\citep{FE85,FH96,SP07}. PNe, on the other hand, are characterized by
narrow emission lines arising from the low velocity expanding outer
shells \citep{BF02,GO09}. Both classes of objects have been
extensively studied by several authors, although just a few are so
dusty that they were initially discovered only at mid-infrared
wavelengths. To distinguish between these two possibilities, further
spectroscopy on OL17 was needed.

In this research note, we present the original IRS spectrum
(\S~\ref{s_irs}) as well as follow-up VLT/X-shooter spectra obtained as
part of the instrument science verification phase (\S~\ref{s_xshoot})
and report on our identification of this object as a PN
(\S~\ref{s_res}).

\section{{\it Spitzer}/IRS data}
\label{s_irs}

\begin{figure}[t]
\begin{center}
\includegraphics[width=\columnwidth]{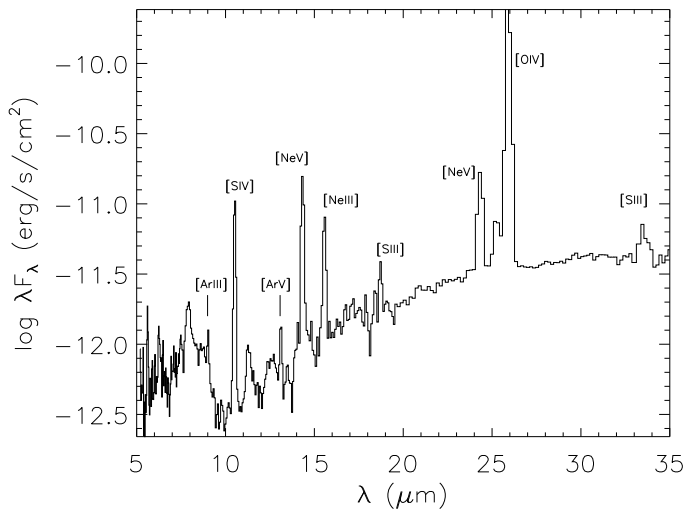}
\end{center}
\caption{\label{irspn} IRS spectrum of OL17, a PN/SNR candidate found
  among the sample from \citet{OL10}. The [O {\sc iv}] line (25.89
  $\mu$m) clearly dominates the spectrum. }
\end{figure}

The IRS spectrum of OL17 ($\alpha_{J2000}$=18$^h$28$^m$27.2$^s$,
$\delta_{J2000}$=+00$^d$44$^m$45$^s$) is presented in Figure
\ref{irspn}. The data were obtained as part of the SST Cycle 3 program
(GO3 30223, PI: Pontoppidan) in the low-resolution module ($R =
\lambda/ \Delta \lambda$ = 50 -- 100; Short Low [SL], 5.2 -- 14.5
$\mu$m and Long Low [LL], 14.0 -- 38.0 $\mu$m). See \citet{OL10} for
further details on the program and the procedures for data reduction.

The spectrum is dominated by strong [O {\sc iv}] emission\footnote{In
  principle, the [O {\sc iv}] may suffer from contamination by [Fe
    {\sc ii}] at the resolution of the LL module, but this is ruled
  out in this case following our conclusion that this object is a PN
  and given that [Fe {\sc ii}] is relatively weak in such objects.} at
25.89 $\mu$m, and is accompanied by other prominent high ionization
lines: [Ar {\sc iii}] (8.99 $\mu$m), [S {\sc iv}] (10.51 $\mu$m), [Ar
  {\sc v}] (13.07 $\mu$m), [Ne {\sc v}] (14.32, 24.32 $\mu$m), [Ne
  {\sc iii}] (15.56 $\mu$m), and [S {\sc iii}] (18.71 $\mu$m, 33.48
$\mu$m). The equivalent width (EW) of these lines, calculated from
Gaussian fits, as well as the lines fluxes are given in Table
\ref{t_irs}. Furthermore, the spectrum shows clear evidence for
polycyclic aromatic hydrocarbon (PAH) emission (most notably at 6.2,
7.7, 11.2, and 12.8 $\mu$m). This emission most likely comes from a
shell of dusty material that surrounds, and is illuminated by, the
central object.

\begin{figure}[t]
\begin{center}
\includegraphics[width=\columnwidth]{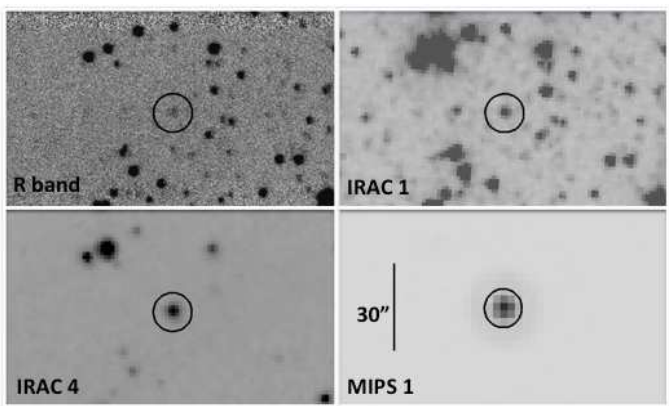}
\end{center}
\caption{\label{sn_image}Optical and infrared images showing the
  INT/WFC $R$-band ({\it top-left}), IRAC1 ({\it top-right}), IRAC4
  ({\it bottom-left}) and MIPS1 ({\it bottom-right}). All three images
  display the same field of view and a length scale of 30$\arcsec$ is
  shown for reference. }
\end{figure}

Because OL17 lies along the line of sight towards the Serpens
Molecular Cloud, it is appreciably extincted due to foreground
material ($A_V \sim$ 6.0 mag). It has weak detections in both the
$R$-band and IRAC1 band (3.5 $\mu$m), but is very prominent in the
MIPS1 band (24 $\mu$m). It has a 3--24 $\mu$m slope ($\frac{d
  \log(\lambda F_\lambda)}{d \log(\lambda)}$) of $\approx$1.8. As
shown in Fig. \ref{irspn}, the MIPS1 band is dominated by the very
strong [O {\sc iv}] emission line. This explains why this object
appeared so red in the infrared color-magnitude diagrams, leading to
the original misclassification of the source as an YSO. The optical
$R$-band image (taken with the Wide Field Camera (WFC) on the Isaac
Newton Telescope (INT, \citealt{SP10}), the IRAC1 \citep{HA06} and
MIPS1 \citep{HB07} images are shown in Figure \ref{sn_image} (see
Table \ref{t_phot} for the photometry in different bands).

\section{X-shooter data}
\label{s_xshoot}

X-shooter \citep{DO06} is a unique multi-wavelength (300-2480 nm)
medium resolution ($R = 4000-14000$) spectrograph on the Cassegrain
focus of UT2 on the Very Large Telescope (VLT). It consists of 3
independent arms that give simultaneous spectra long-ward of the
atmospheric cutoff in the UV (`UVB' arm), optical (`VIS' arm) and
near-infrared (`NIR' arm). The data\footnote{Program ID 60.A-9416(A)}
presented here were obtained as part of the X-shooter 2nd science
verification phase (SV2), during September--October 2009
(60.A-9416(A), PI: Oliveira). The total amount of time awarded was 1
hr, during which integrations of 2760, 2560 and 2880s were obtained
with the UVB, VIS and NIR arms, respectively. The observations were
performed using the $1\arcsec\times12\arcsec$ slit in
`nodding-on-slit' mode to perform adequate background subtraction in
the near-infrared. The spectral resolving powers achieved in this
setup were $\approx$5100 in UVB and NIR and $\approx$8800 in VIS.  The
raw science data consist of images containing highly curved,
multi-order, two-dimensional spectra that need to be rectified,
wavelength calibrated and merged using an input model configuration
that describes the X-shooter spectral format. Data reduction was
performed using an internal release of the X-shooter data calibration
pipeline version 0.9.4 (Andrea Modigliani, private communication),
together with the `com4' reference calibration data.

\begin{figure*}[t]
\begin{center}
\includegraphics[width=0.8\textwidth]{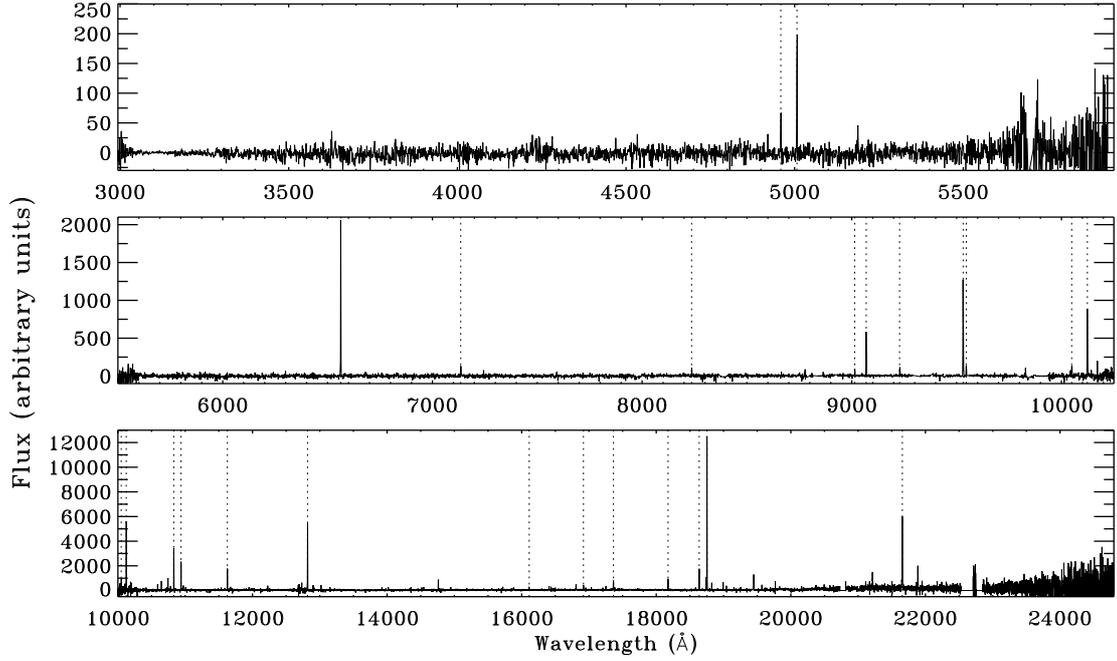}
\end{center}
\caption{\label{xshoot} X-shooter spectra of OL17. Panels show the
  one-dimensional spectra extracted from the UV/blue ({\it top}), the
  visible ({\it middle}), and the near-infrared ({\it bottom})
  arms. Dotted lines mark the major emission lines tabulated in Table
  \ref{t_xs}.}
\end{figure*}

The extracted one-dimensional spectra are shown in Figure
\ref{xshoot}, and show many strong, narrow emission lines especially
in the optical and NIR. These spectra were extracted using an aperture
with a length of $\sim5\arcsec$ in order to include most of the
extended emission seen along the slit (see top panel of Figure
\ref{ha} for the two-dimensional spectrum). Since OL17 is seen through
a dense molecular cloud, it is very faint. These exposures therefore
allowed the detection of the brightest emission lines, but not the
continuum. Although the spectra were not flux calibrated, the analysis
below makes use of line ratios involving lines that are closely spaced
in wavelength, and not affected by strong telluric absorption bands.

The main lines identified, and their EWs are listed in Table
\ref{t_xs}. The EWs were calculated by fitting the lines and a
possible continuum using a Gaussian profile and should be considered
(3$\sigma$) lower limits, as the continuum was not detected at a
sufficient signal-to-noise in these observations. It is interesting to
note that the faintness of the continuum as inferred from the optical
spectrum implies that the object detected in the $R$-band image shown
in the top-left panel of Fig. \ref{sn_image} must consist almost entirely of
H$\alpha$ line emission.

\section{Results}
\label{s_res}

\begin{figure}[t]
\begin{center}
\includegraphics[width=\columnwidth]{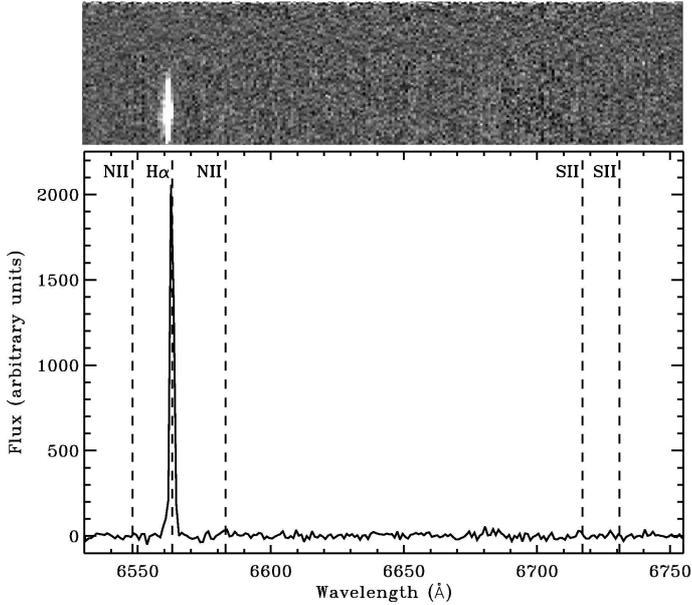}
\end{center}
\caption{\label{ha} A zoom-in of the optical spectrum covering
  H$\alpha$ and the [N {\sc ii}] and [S {\sc ii}] line doublets. The
  weakness of [N {\sc ii}] and [S {\sc ii}] relative to H$\alpha$
  indicate that this object is a PN and not a SNR (see \S~\ref{s_res}
  for details). The top panel shows the two-dimensional spectrum at
  corresponding wavelengths (white pixels indicate large, positive
  pixel values). The slit height measures 12\arcsec, while the spatial
  extent of the resolved H$\alpha$ is about 6\arcsec.}
\end{figure}

According to \citet{FE85}, the main quantitative criterion used to
distinguish SNRs from PNe is the strength of the [S {\sc ii}]
($\lambda\lambda$6717, 6731 \AA) emission lines relative to that of
H$\alpha$. That is, [S {\sc ii}]/H$\alpha > 0.4$ for typical SNRs
\citep{MF97,BL04}. Furthermore, SNRs typically have very strong [N
  {\sc ii}] ($\lambda\lambda$6548, 6583 \AA) emission. Figure \ref{ha}
shows a zoom of the X-shooter/VIS spectrum in the wavelength range
covering H$\alpha$ and the [N {\sc ii}] and [S {\sc ii}] doublets as
labelled. It is clearly seen that the [N {\sc ii}] and [S {\sc ii}]
lines are not detected, while H$\alpha$ is very strong. Upper limits
for the undetected lines were derived, yielding a flux ratio of [N
  {\sc ii}]/H$\alpha \sim$ [S {\sc ii}]/H$\alpha \le 0.06$
(3$\sigma$). This value is well below the value of 0.4 quoted above as
the lower threshold for SNRs.

\begin{figure}[t]
\begin{center}
\includegraphics[width=\columnwidth]{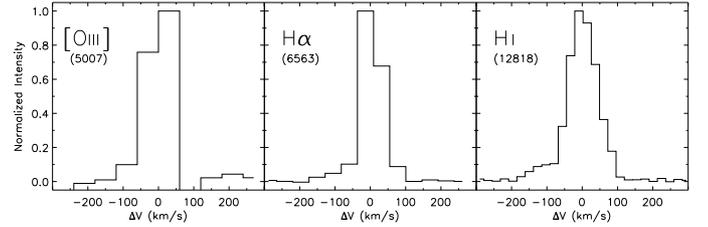}
\end{center}
\caption{\label{kms} A sample of the narrow emission lines present in
  the X-shooter spectra: [O {\sc iii}] ({\it left}) at 5007 \AA,
  H$\alpha$ ({\it middle}) and H {\sc i} (Pa$\beta$, {\it right}) at
  12818 \AA.}
\end{figure}

Furthermore, SNRs have broad, high velocity ($\Delta v \ge$ 300 km
s$^{-1}$) emission lines, while PNe show much narrower lines. This
criterion was used by \citet{FM10} to classify the Spitzer source
J222557$+$601148, previously believed to be a young galactic SNR
\citep{MO06}, as a PN. As it can be seen in Figure \ref{kms} for a
selection of lines in all 3 X-shooter arms, the typical FWHM for lines
seen in these data are $<$15 -- 74 km s$^{-1}$ (deconvolved for
instrumental resolutions of FWHM $\sim$59 in UVB/NIR and 34 km
s$^{-1}$ in VIS). The low velocity of the lines, combined with the low
ratio of [S {\sc ii}]/H$\alpha \le 0.06$ allow the conclusion that OL17
is a PN and not a SNR.

The electron density (n$_e$) can be derived from the ratio between the
[S {\sc iii}] lines at 33.48 and 18.71 $\mu$m in the IRS spectrum,
using the method of \citet{HE82}. The ratio of [S {\sc
    iii}]$\lambda$33.48/18.71$\sim1.9$ yields an electron density of
$\sim$1--100 cm$^{-3}$ \citep{RU89}, where the large range is due to
the fact that we are in the regime where the line ratio is least
sensitive to the density. However, we should note that the line at
33.47$\mu$m appears to be twice as broad as that at 18.71$\mu$m
(possibly due to a data artifact). If we conservatively estimate that
$\sim50$\% of the flux in this line is indeed due to contamination,
the [S {\sc iii}] 33.47/18.71 ratio would yield a density of order
1000 cm$^{-3}$, more typical of PNe. In any case, we note that unlike
the intensity ratios of [N{\sc ii}]/Ha and [S {\sc ii}]/H$\alpha$,
n$_e$ is a rather poor diagnostic for discriminating between SNRs and
PNe as both span a similarly large range in density \citep{SA77,RL05}.

While the distance to this PN cannot be determined without knowing its
intrinsic size, its angular extent of $\approx$6\arcsec\ (determined
from the extent of H$\alpha$ in the two-dimensional spectrum) and the
large extinction it experiences in the bluer parts of the spectrum
suggest that this PN must lie at a distance of at least several kpc,
i.e., firmly behind the Serpens Molecular Cloud (distance of $\sim$415
pc, \citealt{DZ10}). 

We summarize as follows:

\begin{enumerate}

\item The {\it Spitzer}/IRS spectrum of SSTc2dJ18282720+0044450 from
  \citet{OL10} shows strong high ionization emission lines, compatible
  with both Planetary Nebulae and Supernova Remnants, one of only a
  handful such objects identified in the mid-IR to date.

\item High quality VLT/X-shooter spectra, obtained during the science
  verification run, show a wealth of narrow emission lines in the
  UV/blue, optical and near-IR arm data, illustrating the enormous
  potential of X-shooter as a diagnostic tool for optically faint
  sources discovered at other wavelengths.

\item The optical line ratios are compatible with PNe and not with
  SNRs: the low values measured for the line ratio [S {\sc
      ii}]/H$\alpha$, one of the main diagnostic criteria for
  distinguishing between the two types of objects, allows a
  straightforward classification of this object as a PN.

\item The PN interpretation is confirmed by the relative narrow
  ($\Delta v \sim$8--74 km s$^{-1}$) width of the emission lines
  seen throughout the spectra.

\item This previously unknown PN now has public, high-quality
  spectroscopic data covering a very large bandwidth ranging from 0.3
  to 35 $\mu$m. We encourage the use of these data\footnote{The data
    are publicly available and can be accessed through the data
    archives at ESO and the Spitzer Science Center:
    \\\texttt{http://archive.eso.org/eso/eso\_archive\_main.html}\\ \texttt{http://ssc.spitzer.caltech.edu/spitzerdataarchives/}}
  by the community wishing to further study objects of this class.

\end{enumerate}

\begin{acknowledgements}
  Astrochemistry at Leiden is supported by a Spinoza grant from the
  Netherlands Organization for Scientific Research (NWO) and by the
  Netherlands Research School for Astronomy (NOVA) grants. The authors
  would like to express their gratitude to Bram Venemans and Brent
  Groves for help with the X-Shooter data reduction and electron
  densities, respectively.
\end{acknowledgements}

\begin{table}
\caption[ ]{\label{t_irs} Emission lines identified in the IRS
  spectrum of OL17 (Figure \ref{irspn}).}
\begin{center}
\scriptsize
\begin{tabular}{cccc}  
\hline
\hline
$\lambda$ & Line & EW   & Flux  \\
 ($\mu$m) & &  ($\mu$m) & (erg/s/cm$^2$) \\
\hline
 8.99  &  [Ar {\sc iii}] &  0.08 & 6.15 $\times$ 10$^{-15}$ \\
10.51  &  [S {\sc iv}]   &  2.91 & 1.26 $\times$ 10$^{-13}$ \\
13.07  &  [Ar {\sc v}]   &  0.14 & 6.89 $\times$ 10$^{-15}$ \\
14.32  &  [Ne {\sc v}]   &  3.23 & 1.88 $\times$ 10$^{-13}$ \\
15.56  &  [Ne {\sc iii}] &  1.11 & 8.30 $\times$ 10$^{-14}$ \\
18.71  &  [S {\sc iii}]  &  0.23 & 2.03 $\times$ 10$^{-14}$ \\
24.32  &  [Ne {\sc v}]   &  1.45 & 1.92 $\times$ 10$^{-13}$ \\
25.89  &  [O {\sc iv}]   & 13.64 & 2.81 $\times$ 10$^{-12}$ \\
33.48  &  [S {\sc iii}]  &  0.31 & 3.80 $\times$ 10$^{-14}$ \\
\hline
\end{tabular}
\end{center}
\end{table}

\begin{table}
\caption[ ]{\label{t_phot} Optical R-band magnitude and {\it Spitzer}
  IRAC/MIPS fluxes of OL17 (see also Figure \ref{sn_image}).}
\begin{center}
\scriptsize
\begin{tabular}{cccccc}  
\hline
\hline
R$^\dag$ & 3.6 $\mu$m & 4.5 $\mu$m & 5.8 $\mu$m & 8.0 $\mu$m & 24.0 $\mu$m\\
 (mag) &  (mJy)  &  (mJy)  &  (mJy)  &  (mJy)  &  (mJy) \\
\hline
 22.8 & 0.28 & 0.64 & 0.80 & 2.29 & 69.60 \\
\hline
\end{tabular}
\end{center}
$^\dagger$ \footnotesize{From \citealt{SP10}.}
\end{table}

\begin{table}
  \caption[ ]{\label{t_xs} Major emission lines identified in the
    X-shooter spectra (Figure \ref{xshoot}). Equivalent widths (EW)
    given indicate (3$\sigma$) lower limits, due to the faintness of
    the continuum.}
\begin{center}
\scriptsize
\begin{tabular}{rcc|rcc}  
\hline
\hline
\multicolumn{1}{c}{$\lambda$} & Line & EW &\multicolumn{1}{c}{$\lambda$} & Line & EW \\
\multicolumn{1}{c}{(\AA)} & & (\AA) & \multicolumn{1}{c}{(\AA)} & & (\AA)\\
\hline
 4959 & [O {\sc iii}]  &   1.4 & 10831 & He {\sc i}     & 196.1 \\
 5007 & [O {\sc iii}]  &   5.1 & 10938 & H {\sc i}      & 125.7 \\
 6563 & H$\alpha$      &  47.1 & 11627 & He {\sc ii}    & 100.4 \\
 7135 & [Ar {\sc iii}] &   2.6 & 12818 & H {\sc i}      & 290.5 \\
 8236 & He {\sc ii}    &   2.9 & 16112 & O {\sc i}      &  21.9 \\
 9014 & H {\sc i}      &   2.4 & 16918 & Al {\sc ii}    &  25.6 \\
 9069 & [S {\sc iii}]  &  18.8 & 17364 & Ne {\sc ii}    &  66.0 \\
 9229 & H {\sc i}      &   5.5 & 18176 & Ne {\sc ii}    &  88.1 \\
 9531 & [S {\sc iii}]  &  43.4 & 18638 & Ne {\sc ii}    & 183.4 \\
 9546 & H {\sc i}      &  22.2 & 18754 & N {\sc ii}     & 564.4 \\
10049 & H {\sc i}      &   5.4 & 21657 & Br $\gamma$    & 597.1 \\
10123 & He {\sc ii}    & 228.5 & \\
\hline
\end{tabular}
\end{center}
\end{table}

\end{document}